# Acoustic Impedance Prediction Using an Attention-Based Dual-Branch Double-Inversion Network

Wen feng[1], Yong Li[1,2*], Yingtian Liu[1], Huating Li[1]

*1 Key Laboratory of Earth Exploitation and Information Technology of Ministry of Education, Chengdu University of Technology, China. E-mail: fengwen@stu.cdut.edu.cn*

*2 State Key Lab of Oil and Gas Reservoir Geology and Exploitation, Chengdu University of Technology, China. E-mail: liyongo7@cdut.edu.cn (corresponding author); liuyingtian@stu.cdut.edu.cn; 2390328914@qq.com*

**Abstract** Seismic impedance inversion is a widely used technique for reservoir characterization. Accurate, high-resolution seismic impedance data form the foundation for subsequent reservoir interpretation. Deep learning methods have demonstrated significant potential in seismic impedance inversion. Traditional single semi-supervised networks, which directly input original seismic logging data, struggle to capture high-frequency weak signals. This limitation leads to low-resolution inversion results with inadequate accuracy and stability. Moreover, seismic wavelet uncertainty further constrains the application of these methods to real seismic data. To address these challenges, we propose ADDIN-I: an Attention-based Dual-branch Double-Inversion Network for Impedance prediction. ADDIN-I's dual-branch architecture overcomes the limitations of single-branch semi-supervised networks and improves the extraction of high-frequency weak signal features in sequence modeling. The network incorporates an attention mechanism to further enhance its feature extraction capabilities. To adapt the method for real seismic data applications, a deep learning forward operator is employed to fit the wavelet adaptively. ADDIN-I demonstrates excellent performance in both synthetic and real data applications.

## 1. Introduction

Seismic impedance inversion plays a critical role in oil and gas exploration. Acoustic impedance, a key parameter reflecting strata characteristics such as lithology, porosity, and fluid saturation, enables the prediction of subsurface physical properties (Doyen, 2007; Yin and Zhang, 2014; Shi et al., 2014). The relationship between seismic data and rock properties is non-linear and affected by factors like measurement errors, processing defects, and noise. To enhance result credibility, prior information and assumptions are essential for constraining inversion outcomes (Wu et al., 2021). Traditional seismic inversion techniques use regularization constraints or structural guidance to solve inversion challenges (Zhang and Castagna, 2011; Zhang et al., 2013). This approach often necessitates tailored inversion techniques for specific regions, leading to poor model generalization. Additionally, traditional methods require an a priori mathematical-physical model linking model parameters to measured data. Complex geological structures make accurate modeling difficult, often resulting in substantial differences between inversion results and actual data (Wang et al., 2021). Common inversion techniques include recursive (Cooke and Schneider, 1983), simultaneous (Hampson et al., 2005), elastic impedance (Verwest et al., 2000), and AVO inversion (Buland and More, 2003). Traditional inversion approaches face challenges including complex wave equation analysis, lengthy simulations, and high computational demands. Researchers are leveraging artificial intelligence to overcome these challenges in geophysics, applying it to fault interpretation (Wu et al., 2018), seismic data denoising (Saad and Chan, 2020), stratigraphic horizon estimation (Geng et al., 2020), and seismic inversion (Chen et al., 2021; Das et al., 2019; Zhang et al., 2021).

Deep learning methods model non-linear inversion processes using multiple processing layers and appropriate prior knowledge, achieving impedance inversion directly and bypassing the complex mathematical problems inherent in traditional methods (LeCun et al., 2015; Chai et al., 2021). Kim and Nakata (2018) compared traditional and deep learning inversion methods, concluding that deep learning approaches offer superior accuracy. Das et al. (2019) applied convolutional neural networks (CNNs) to impedance inversion, achieving an 82% correlation with impedance logs. Alfarraj and AlRegib (2019) inverted impedance by modeling impedance and seismic data as time series. Biswas et al. (2019) proposed a novel approach combining CNN-based unsupervised learning with physical constraints for inversion problems. Wang et al. (2020) developed a closed-loop CNN impedance inversion method that extracts features from both labeled and unlabeled data, reducing the CNN's reliance on large volumes of labeled data. Meng et al. (2021) introduced a GAN-based semi-supervised deep learning method that outperformed traditional deep learning approaches in impedance inversion. Wu et al. (2021) enhanced inversion performance by replacing 1D networks with 2D CNNs and incorporating initial impedance model constraints. Zhang et al. (2022) investigated the impact of network hyperparameters and structures on inversion performance, designing a multi-scale CNN that effectively reconstructs high-frequency information in estimated impedance. Wang et al. (2022) proposed a novel seismic impedance inversion method using cycle-consistent generative adversarial networks (cycle GANs). Song et al. (2022) developed a spatiotemporal sequence residual modeling network to address the challenge of deriving subsurface impedance from zero-offset seismic data and initial impedance models. Sang et al.

(2023) demonstrated promising porosity prediction results using bidirectional recurrent units in a well-seismic joint approach. In conclusion, acoustic impedance inversion methods have evolved from traditional mathematical models to advanced deep learning approaches, markedly improving inversion accuracy and efficiency.

Traditional single semi-supervised networks have notable limitations in seismic data processing. These networks directly process raw seismic data but struggle to capture high-frequency weak signals. This limitation leads to low-resolution inversions and compromises stability and generalization (Zhu et al., 2022). Moreover, the uncertainty in seismic wavelets further limits the application of these networks to real seismic data.

We propose an acoustic impedance prediction method using an attention-enhanced dual-branch double inversion network to address these challenges. The dual-branch architecture enhances feature extraction in sequence modeling through two approaches: 1) Temporal Convolutional Networks (TCN) with varying dilation factors create multi-scale convolutional receptive fields, extracting rich features. 2) Bidirectional Gated Recurrent Units (BI-GRU) perform sequence modeling. These complementary methods effectively address the challenge of extracting high-frequency weak signals. We define this two-channel feature extraction approach as "dual-branch". Next, we integrate attention mechanisms to facilitate automatic learning of feature relationships. This mechanism calculates weights to highlight key input features, improving both prediction accuracy and network stability. We construct a deep learning forward operator to address unknown seismic wavelets, facilitating the method's application to real seismic exploration. We term this process "double inversion" as the deep learning forward operator and inversion network are trained and updated simultaneously. We also propose an improved semi-supervised learning strategy for network training using partitioned data. This strategy effectively leverages limited labeled well log data and abundant unlabeled seismic data, enhancing the model's predictive performance. Compared to traditional methods, our approach improves both data utilization efficiency and prediction accuracy.

The structure of this paper is as follows: Section 2 presents the theoretical foundation; Section 3 compares the inversion results of various methods using model data and demonstrates their application to real seismic data; Section 4 concludes the paper.

## 2. Methodology

### 2.1. Attention Block

Inspired by the human visual system, the attention mechanism focuses on key elements through weighted calculations, rather than processing the entire scene, thus enhancing visualization (Lu et al., 2023). We incorporate attention modules into our inversion network to enhance input data feature capture. Attention mechanisms, implemented through task-specific neural networks, have demonstrated effectiveness across various applications (Ran et al., 2019). The attention module enables automatic learning of feature relationships, emphasizing information-rich features and stabilizing hidden data structure extraction (Hu et al., 2018). Figure 1(a) illustrates the structure of this module. The module comprises two Dense layers, two activation function layers, and a multiplication layer, represented as:

$$X^{l-1} = M_{channel}(H^{l-1}) \otimes H^{l-1} \tag{1}$$

$$M_{channel}(H^{l-1}) = \sigma_S(Den(\sigma_R(Den(H^{l-1})))) \tag{2}$$

where $\otimes$ denotes element-wise multiplication; $X^{l-1}$ represents the output features of the $L-1$ th attention block; Den indicates a fully connected layer; $\sigma_S$ denotes the non-linear Sigmoid activation function; and $\sigma_R$ denotes the non-linear ReLU activation function.

### 2.2 Bidirectional gated recurrent units (Bi-GRU)

Traditional deep learning networks often fail to capture the nonlinear relationships between seismic data and rock physical parameters, instead only fitting relationships between discrete data points. As a result, these networks frequently miss the partial correlations present in continuous seismic and well log data sequences (Alfarra and AlRegib, 2019). The Gated Recurrent Unit (GRU), an improved Recurrent Neural Network (RNN) unit for sequential data processing, was introduced by Cho et al. (2014). Figure 1(b) illustrates its basic structure. GRU improves upon traditional RNNs by introducing two gating mechanisms: reset and update gates. These gates enable intelligent updating or resetting of memory states based on input sequences. This design effectively captures long-term dependencies and mitigates the vanishing or exploding gradient problems common in traditional RNNs processing long sequences (Hochreiter and Schmidhuber, 1997). Bidirectional GRU (Bi-GRU) enhances unidirectional GRU by incorporating a reverse recurrent layer, allowing simultaneous capture of forward and backward contextual information in sequences. Bi-GRU consists of two oppositely directed GRU layers, bidirectionally encoding input sequences to generate forward and backward hidden state sequences. This architecture enables synchronous integration of bidirectional contextual features, improving sequence modeling performance. Seismic and well log data display internal correlations and localized interweaving patterns. At any given time or depth point, seismic and well log responses reflect the composite properties of surrounding rock, creating contextual dependencies. Bi-GRU's bidirectional recurrent structure captures both forward and backward contextual dependencies, comprehensively characterizing long-term dependencies in seismic data. This capability enables theoretical integration of seismic and well log data for reservoir parameter prediction. For a comprehensive understanding of Bi-GRU principles, refer to Cho et al. (2014), Tang et al. (2019), and Liu et al. (2024). This study employs four Bi-GRU units for sequence modeling. Figure 1(e) illustrates their basic structure.

**2.3 Temporal Convolutional Block with Integrated Attention Mechanism**

Temporal Convolutional Networks (TCNs) address the challenge of capturing long-term dependencies in extended sequence data (Bai et al., 2018). TCNs provide an efficient architecture for sequential data processing, outperforming recurrent neural networks in parallelism and computational efficiency (Luo et al., 2023). The core component of a TCN is the temporal convolutional block, comprising two convolutional layers alternating with weight normalization, dropout, and nonlinear activation layers. We introduced dilation factors to adjust the receptive field size of TCN's convolutional layers (Yu and Koltun, 2015). These factors control the spacing between convolutional kernel elements, allowing for receptive field size adjustment during convolutions. This approach extracts richer features, enhancing overall model performance. We also integrated attention layers into the temporal convolutional blocks (Figure 1(c)). These attention layers enable direct information exchange between time steps, improving the network's capacity to model long-term dependencies. This integration enables automatic learning of inter-feature relationships, boosting overall performance. In our study, we constructed the temporal convolutional network using temporal convolutional blocks with integrated attention layers. Figure 1(d) illustrates the structure of this network.

## 2.4 Inversion Network

Seismic data contains rich spatiotemporal information. Accurately capturing its intrinsic sequential dependencies and comprehensive features is crucial for model prediction (Arnold et al., 2019). Our inversion network addresses this challenge through four modules, as shown in Figure 1(f): 1) Bi-GRU sequence modeling, 2) Temporal convolution with integrated attention blocks, 3) Upsampling, and 4) Regression. The Bi-GRU sequence modeling module treats input seismic sequences as time series, computing temporal features from dynamic input trace changes to obtain effective representations. Deep Bi-GRUs typically generate smooth outputs, yielding low-frequency feature information (Alfarraj and AlRegib, 2019). These low-frequency features often contain crucial information, such as lithological variations at formation interfaces, providing valuable sequential prior knowledge for model predictions.

The temporal convolution module uses one-dimensional dilated convolutions with varying dilation factors and integrated attention blocks. This module captures information from both adjacent and distant trace samples, achieving multi-scale feature extraction and yielding richer feature data. The enriched features are then fused with low-frequency features from the other channel, compensating for high-frequency weak signals and producing comprehensive full-bandwidth feature output.

An upsampling layer resolves the resolution mismatch between seismic and well log data. This layer comprises two cascaded deconvolution blocks (De-Conv-Block) (Alaudah et al., 2018). The stride size determines the input data's upsampling factor. Each deconvolution block contains a deconvolution layer, a group normalization layer, and a tanh activation function layer. The regression module maps features from other modules to the target domain, achieving the final inversion objective. This module consists of a bidirectional gated recurrent unit (Bi-GRU) and a fully connected layer (Linear).

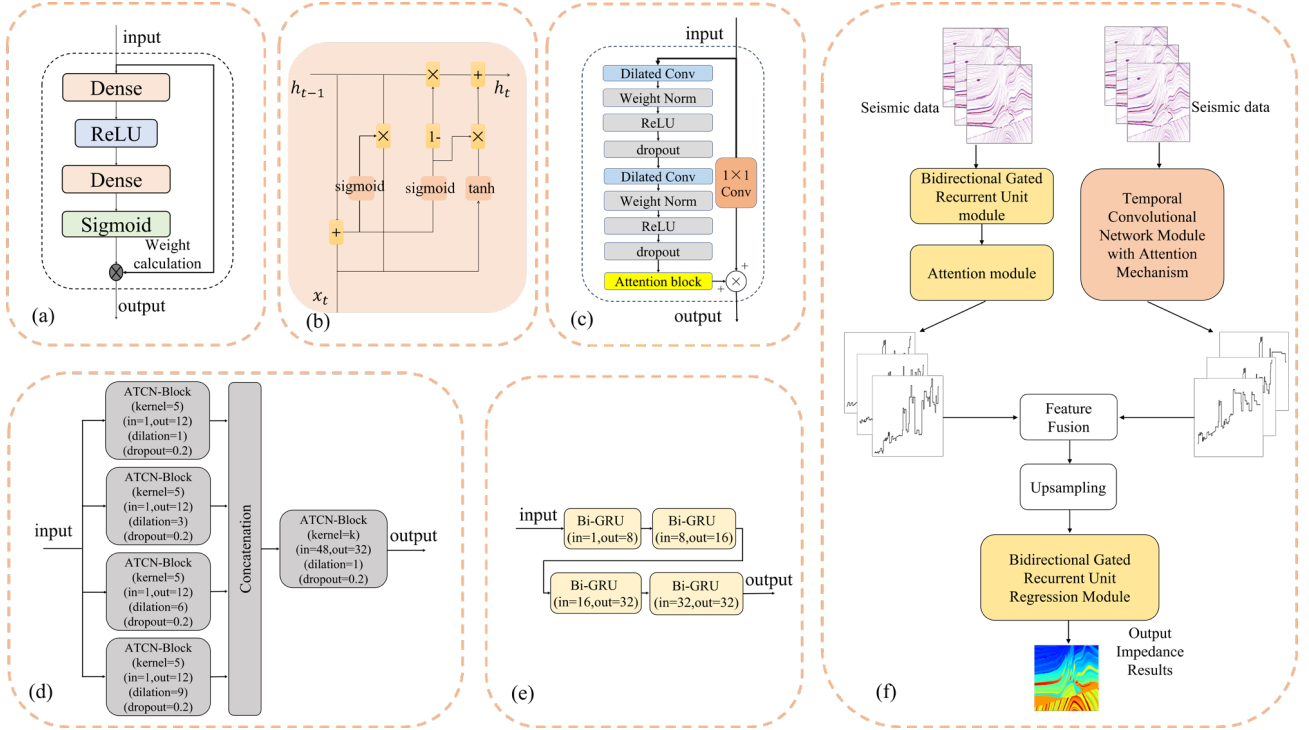

Figure 1. Model Components and Architecture. (a) Attention block structure; (b) GRU structure. ($x_t$: The input vector at time t; $h_t$: Hidden feature vector at time t; $h_{t-1}$: Hidden feature vector at time t-1) ; (c) TCN-Block with integrated attention block (ATCN-Block); (d) Temporal Convolutional Network Architecture; (e) Sequence modeling module structure; (f) Dual-branch Inverse model architecture.

## 2.5 Semi-Supervised Network Training Strategy

Traditional semi-supervised inversion methods use a closed-loop process (Alfarraj and AlRegib, 2019). The process starts by inputting seismic data into an inversion model to predict acoustic impedance. The predicted acoustic impedance then undergoes forward modeling to create synthetic seismic data. A seismic loss is calculated by comparing the synthetic and original input seismic data. Simultaneously, the predicted acoustic impedance is compared with ground truth data from well logs to compute another loss value. These two loss functions guide the iterative optimization of the inversion model's parameters, refining acoustic impedance predictions.

We propose an improved semi-supervised network training strategy based on this framework. Our strategy consists of two main modules (an Inversion Network and a deep learning-based forward modeling operator) and incorporates multiple constraint loss functions. Figure 2 illustrates this structure. Our enhanced method improves inversion accuracy and generalization by differentiating near-well and non-near-well data and using multiple loss functions. The Inversion Network processes both near-well and non-near-well seismic data simultaneously, predicting acoustic impedance. The forward modeling operator then uses predicted and true acoustic impedance to generate synthetic seismic data. Our improved strategy introduces three loss functions: unlabeled seismic, labeled acoustic impedance, and labeled seismic losses. Unlabeled seismic loss compares original and synthetic seismic data in non-near-well regions. Labeled acoustic impedance loss evaluates consistency between predicted impedance and well log data in near-well regions. Labeled seismic loss measures the match between original and synthetic seismic data in near-well regions. Both labeled and unlabeled seismic losses contribute to updating the Inversion Network and forward modeling operator simultaneously. Near-well seismic data, as labeled data, provides more accurate predictions. Thus, loss functions from near-well seismic data are more crucial in updating network parameters. Adjusting weight coefficients in loss functions enhances the influence of near-well seismic data on network training. The multiple loss functions design ensures dual constraints on acoustic impedance estimation, considering both seismic response and rock properties, enhancing model learning capacity.

Traditional semi-supervised acoustic impedance inversion uses forward modeling to synthesize seismic records by convolving seismic wavelets with reflection coefficients. However, difficulties in extracting wavelets from real seismic data have hindered semi-supervised inversion network training. To address this, we propose a deep learning-based forward modeling operator aligned with the inversion network structure. This operator uses an end-to-end strategy to directly predict seismic records, leveraging deep learning advantages in seismic inversion. This approach allows simultaneous training and updating of both operators, enhancing the network's ability to process real seismic and well log data.

This study uses Mean Squared Error (MSE) as the objective function (Chicco et al., 2021). The objective function is expressed as follows:

$$Loss_1 = \frac{1}{n} \sum_{i=1}^{n} (P_i - \hat{P}_i)^2 \quad (3)$$

$$Loss_2 = \frac{1}{n} \sum_{i=1}^{n} (S_i - \hat{S}_i)^2 \quad (4)$$

$$Loss_3 = \frac{1}{n} \sum_{i=1}^{n} (Y_i - \hat{Y}_i)^2 \quad (5)$$

$$L = \lambda_1 Loss_1 + \lambda_2 Loss_2 + \lambda_3 Loss_3 \quad (6)$$

where $i$ represents a seismic trace acquisition point, $L$ is the constructed objective function, Loss1 denotes the loss function between the labeled well acoustic impedance $P_i$ and the predicted acoustic impedance $\hat{P}_i$; Loss2 is the waveform loss function between near-well seismic data $S_i$ and predicted near-well seismic data $\hat{S}_i$; Loss3 is the waveform loss function between non-well seismic data $Y_i$ and predicted non-well seismic data $\hat{Y}_i$; $\lambda_1$, $\lambda_2$, and $\lambda_3$ are the weight coefficients for the respective loss functions. Through empirical testing, the values of $\lambda_1$, $\lambda_2$, and $\lambda_3$ were determined to be 0.7, 0.2, and 0.1, respectively.

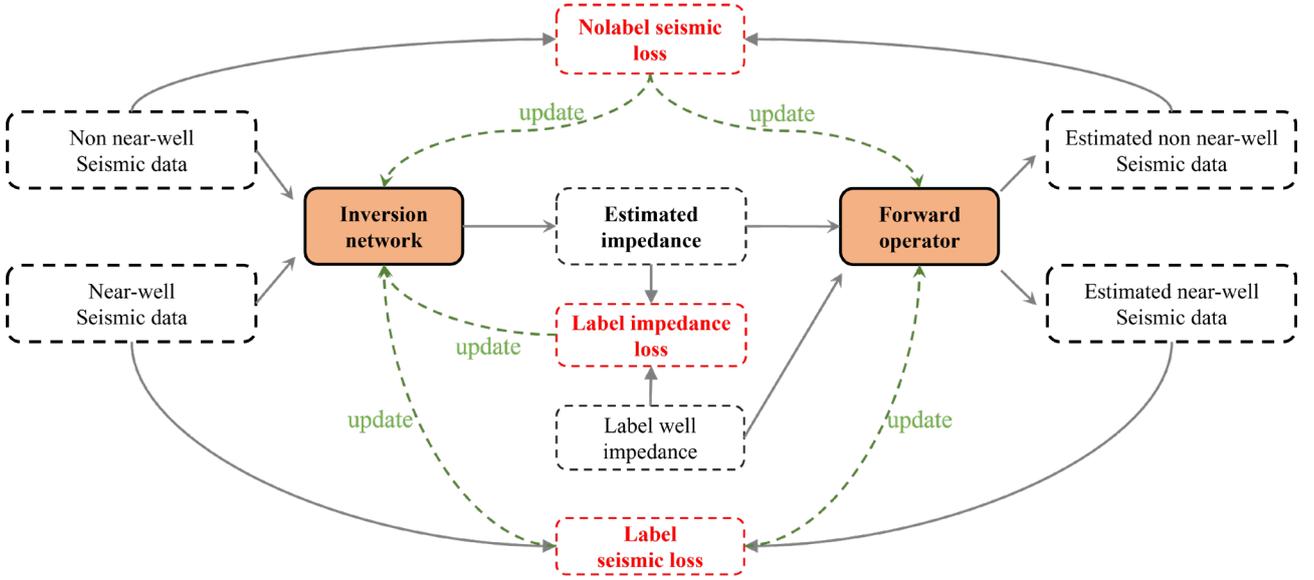

Figure 2. Architecture of the Improved Semi-supervised Network Training.

## 3. Inversion experiment

### 3.1 Data Preprocessing

We validate our proposed method using the Marmousi2 model. Synthetic seismic data were generated by convolving a 40Hz Ricker wavelet with the model. The Marmousi2 model consists of 2500 seismic traces, each with 3500 sample points. Z-score normalization was applied to the data to improve model convergence efficiency and training speed. The Z-score normalization formula is:

$$Z_X = \frac{X - X_{mean}}{X_{std}} \tag{7}$$

$$Z_Y = \frac{Y - Y_{mean}}{Y_{std}} \tag{8}$$

where $X$ represents the original seismic data, and $Y$ denotes the original acoustic impedance; $Z_X$ is the normalized seismic data in the training set, and $Z_Y$ is the normalized acoustic impedance data in the training set; $X_{mean}$ and $Y_{mean}$ are the mean values of seismic data and acoustic impedance data, respectively; $X_{std}$ and $Y_{std}$ are the standard deviations of seismic data and acoustic impedance data, respectively.

The model training set was extracted following the principle of sparse uniformity. Nineteen traces (≤ 1% of total traces) of synthetic seismic records and acoustic impedance were extracted at equal intervals for model training, as illustrated by black curves in Figure 3. For evaluating model generalization, ten traces were extracted at equal intervals from the profile (excluding the training set) as the validation set. The remaining unselected data served as unlabeled data for training the semi-supervised double inversion network.

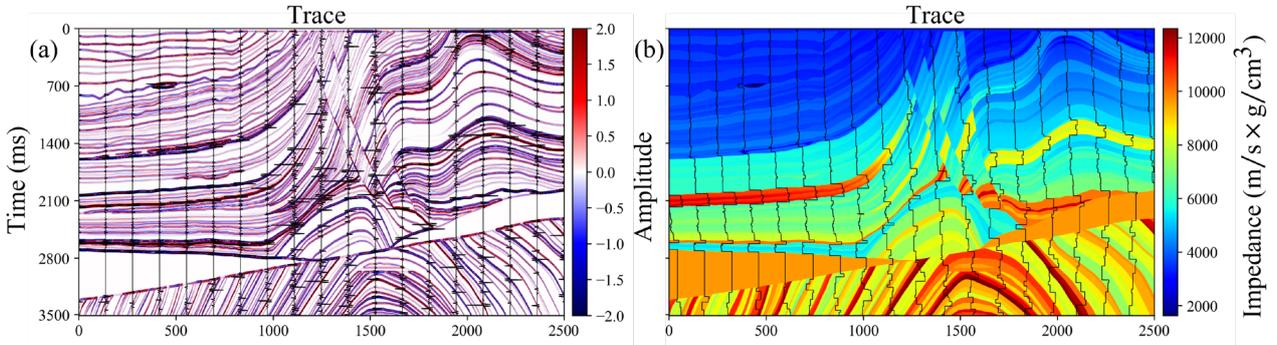

Figure 3. Training Data Gathers. (a) Seismic data; (b) Impedance data.

Model hyperparameters were optimized before training. Dropout regularization with a coefficient of 0.2 was implemented (Srivastava et al., 2014). Adam optimizer was used with an initial learning rate of 0.005 (Kingma and Ba, 2014). The tanh function was chosen for activation (Fan, 2000). The model was trained for 500 epochs with a batch size of 10. Table 1 presents detailed hyperparameter configurations. Experiments were conducted using a computer with an Intel i5 quad-core CPU and a GeForce GTX 2080 Ti GPU.

Table 1. Network Hyperparameter Settings

|  | Hyperparameter type | value |
| --- | --- | --- |
| Basic parameters | epoch | 500 |
|  | Batch size | 10 |
|  | Learning rate | 0.005 |
|  | Dropout | 0.2 |
|  | Weight coefficient of the loss function | 0.7、0.2、0.1 |
| TCN-Block | Kernel size | 5 |
|  | Dilation | [1,3,6,9] |
|  | Number of channels | [12,12,12,12,32] |
| Bi-GRU | Number of channels | [8,16,32,32,8] |

| | Layers | 4 |
|---|---|---|
| | Kernel size | 4 |
| Conv-Block | Number of channels | [16,16] |
| | Stride | 2 |
| | Kernel size | 4 |
| De-Conv-Block | Number of channels | [16,16] |
| | Stride | 2 |

## 3.2 Evaluation Metrics

We evaluate model performance using two standard metrics: the coefficient of determination (R²) and the Pearson Correlation Coefficient (PCC). R² measures the goodness of fit between variables and is calculated as:

$$R^2(y,\hat{y}) = 1 - \frac{\sum_{i=1}^{N}(y_i - \hat{y}_i)^2}{\sum_{i=1}^{N}(y_i - \mu_i)^2} \tag{9}$$

where R² ranges from 0 to 1, with higher values indicating better model fit. The PCC measures the linear correlation between two variables, calculated as:

$$PCC(y,\hat{y}) = \frac{1}{N} \frac{\sum_{i=1}^{N}(y_i - \mu_y)(\hat{y}_i - \mu_{\hat{y}})}{\sqrt{\sum_{i=1}^{N}(y_i - \mu_y)^2} \sqrt{\sum_{i=1}^{N}(\hat{y}_i - \mu_{\hat{y}})^2}} \tag{10}$$

where $\mu_y$ and $\mu_{\hat{y}}$ represent the mean values of the input parameters and predicted values, respectively. The PCC ranges from -1 to 1, with positive values indicating positive correlation between variables, negative values indicating negative correlation, and zero indicating no correlation.

## 3.3 Model training results

This study evaluated the impact of various elements on acoustic impedance inversion using the Marmousi2 model dataset. Key components assessed include attention mechanisms, dual-branch structures, double inversion strategies, and improved semi-supervised training strategies. Various methods were compared, each constructed with modules using identical parameters. The study developed five methods: 1. Attention Mechanism-based Dual-Branch Double-Inversion Network (ADDIN-I) 2. Non-Attention Mechanism Dual-Branch Double-Inversion Network (DDIN-I) 3. Attention Mechanism-based Dual-Branch Single-Inversion Network (ADSIN-T) 4. Non-Attention Mechanism Dual-Branch Single-Inversion Network (DSIN-T) 5. Non-Attention Mechanism Single-Branch Single-Inversion Network (SSIN-T). ADDIN-I and DDIN-I were trained using an improved semi-supervised strategy, whereas the other methods employed traditional semi-supervised approaches. Dual-branch networks comprised Bi-GRU and TCN, following the previously described structure. In contrast, the single-branch network used only Bi-GRU. Figure 4 illustrates the impedance inversion profiles for different methods, while Table 2 presents their quantitative inversion results.

Table 2. Quantitative Comparison of Results from Different Methods.

| Methods | $R^2$ | PCC | MSE |
|---|---|---|---|
| SSIN-T | 0.9420 | 0.9801 | 0.0683 |
| DSIN-T | 0.9707 | 0.9927 | 0.0296 |
| ADSIN-T | 0.9845 | 0.9944 | 0.0164 |
| DDIN-I | 0.9728 | 0.9929 | 0.0234 |
| ADDIN-I | **0.9851** | **0.9952** | **0.0160** |

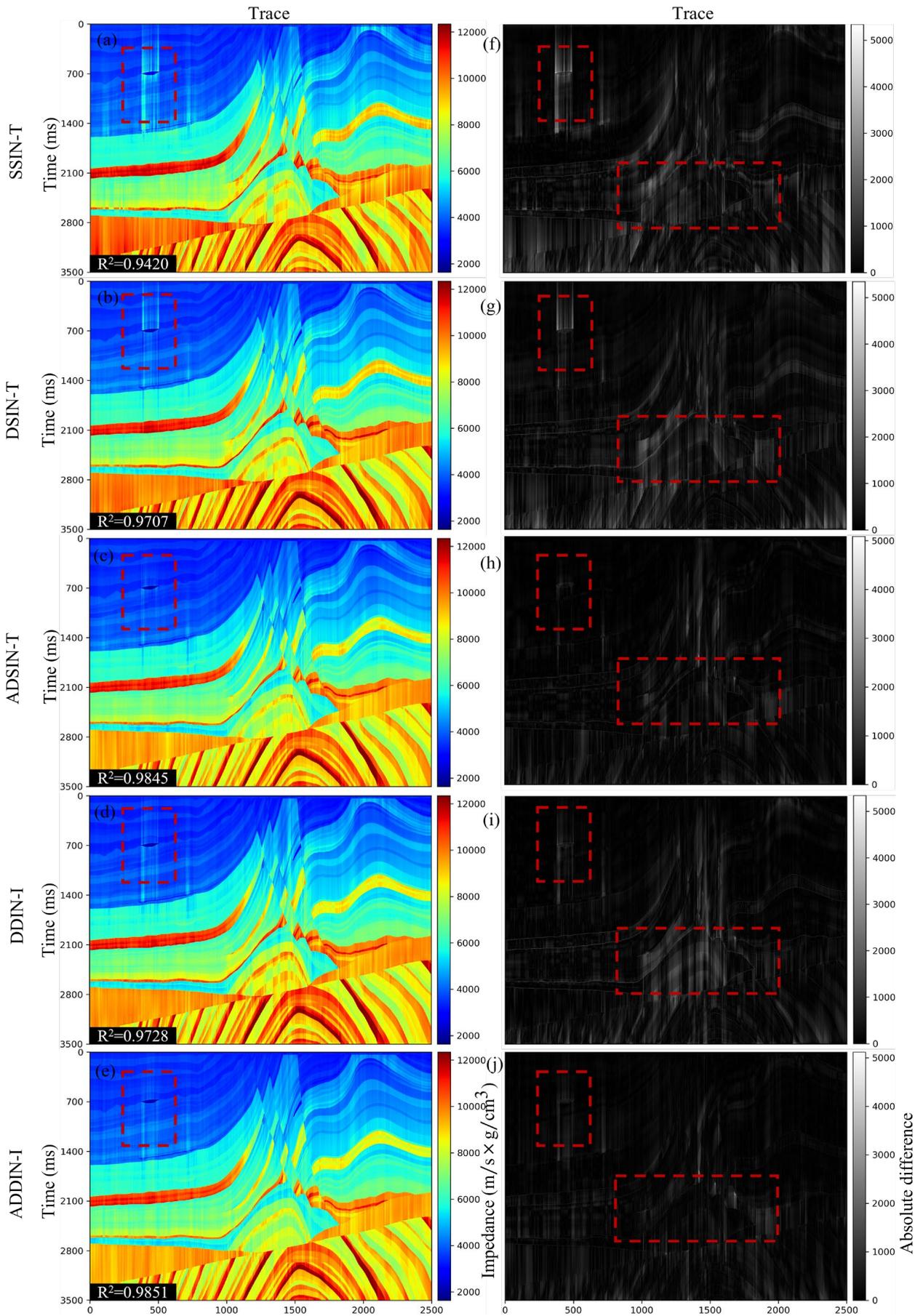

Figure 4. The first row shows the effect plots of the SSIN-T method; the second row shows the effect plots of the DSIN-T method; the third row shows the effect plots of the ADSIN-T method; the fourth row shows the effect plots of the DDIN-I method; the fifth row shows the effect plots of the ADDIN-I method. Figures (a), (b), (c), (d), and (e) represent the inversion results, while figures (f), (g), (h), (i), and (j) show the residual plots (absolute difference between true and predicted acoustic impedance).

Figure 4 presents the inversion results for different methods. The first column shows the inverted profiles, while the second column displays the residual profiles. Each row represents the inversion results from a different method. Figures 4(e) and 4(j) show the predicted impedance and residual profiles, respectively, using the proposed ADDIN-I method. The ADDIN-I inversion results clearly reveal high-resolution structural features of the strata. The ADDIN-I method effectively identifies diverse geological features. It clearly delineates structural elements (fault planes, unconformity surfaces), stratigraphic details (thin layer impedance boundaries), and reservoir characteristics (gas-bearing and oil-bearing sandstone channels). The results show low residuals and high agreement with true impedance profiles. ADDIN-I achieves the highest correlation (0.9851) and Pearson (0.9952) coefficients among all methods. In contrast, SSIN-T yields inferior results, with significant residuals at geological structure discontinuities. SSIN-T has the lowest correlation (0.9420) and Pearson (0.9801) coefficients among all methods. The dual-branch network structure effectively captures high-frequency weak signals that single-branch structures struggle to identify. This enhanced feature extraction leads to more accurate inversion results.

The red-boxed areas in Figures 4(j) and 4(i) reveal significant geological structure changes in the gas-bearing sandstone region. The ADDIN-I method shows lower inversion residuals compared to DDIN-I in this area, suggesting superior performance. Correlation coefficient ($R^2$) analysis further supports the effectiveness of the attention mechanism. ADDIN-I achieves an $R^2$ of 0.9851, while DDIN-I reaches 0.9728. These results confirm the attention mechanism's positive impact on enhancing the predictive capabilities of Bi-GRU and TCN. The ADSIN-T method similarly outperforms DSIN-T in inversion performance. Collectively, these results demonstrate that the attention mechanism enhances model performance in inverting complex geological structures.

From a model data perspective, deep learning operators can asymptotically approach but not completely match synthetic seismic data generated by traditional convolutional forward operators. Under identical conditions, double-inversion networks should theoretically not outperform single-inversion networks. Comparison of inversion results reveals that, contrary to theoretical expectations, ADDIN-I outperforms ADSIN-T. This is evidenced by their respective correlation coefficients: 0.9851 for ADDIN-I and 0.9845 for ADSIN-T. This superior performance can be attributed to the improved semi-supervised network training strategy used in the inversion process. The strategy enhances the influence of labeled data by separating it from unlabeled data and adjusting coefficients, thereby maximizing the use of labeled information. This improved semi-supervised network training strategy proves superior to traditional semi-supervised methods. The deep learning forward operator, based on the inversion network, was verified using supervised learning methods. It achieved a correlation coefficient of 0.9736 when compared to real seismic data. Figure 5 displays the forward seismic data profiles and their residuals. Figure 6 compares the results of the 625th, 1250th, and 1875th traces. The predicted seismic data exhibits three key features that closely match the actual seismic records: high stratigraphic continuity, clear interface resolution, and high fidelity of waveform characteristics. These results demonstrate the deep learning forward operator's excellent performance and indirectly validate the effectiveness of the improved semi-supervised network training strategy.

Figure 7 compares the inverted acoustic impedance to true values for traces 625, 1250, and 1875 across five methods. The figure employs color-coded curves to represent different methods: red for the proposed ADDIN-I method, blue for DDIN-I, purple for ADSIN-T, orange for DSIN-T, and green for SSIN-T. A light gray area denotes one standard deviation range of the true impedance values. Yellow arrows highlight regions of significant geological changes, where the SSIN-T method (green curve) notably underperforms compared to other methods. This indicates that Bi-GRU alone is less effective than the dual-branch Bi-GRU and TCN combination in capturing long-term dependencies and extracting seismic data features. Conversely, the ADDIN-I method (red curve) demonstrates the highest consistency with true data in these critical regions. This result underscores ADDIN-I's superiority in inverting subsurface medium models, producing results that more accurately reflect true geological conditions.

Figure 8 presents scatter plots comparing the inversion performance of each method. We sampled 40 traces uniformly from each method's inversion profile, with 200 data points per trace, yielding 8000 data points per method. The scatter plots show inverted impedance values on the horizontal axis and true impedance values on the vertical axis. A black diagonal line serves as the ideal prediction reference, while a light red area indicates one standard deviation range of the true impedance values. The ADDIN-I method, proposed in this study, demonstrates superior performance with more concentrated data points and the highest consistency with true values compared to other methods. This result underscores the superior inversion accuracy of the ADDIN-I method.

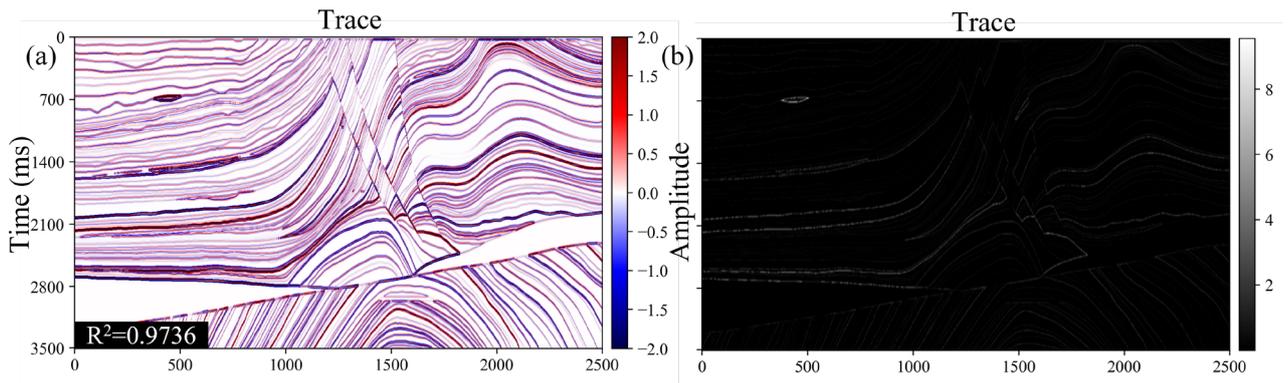

Figure 5. (a) Seismic data profile synthesized by deep learning forward operator; (b) Residual plot.

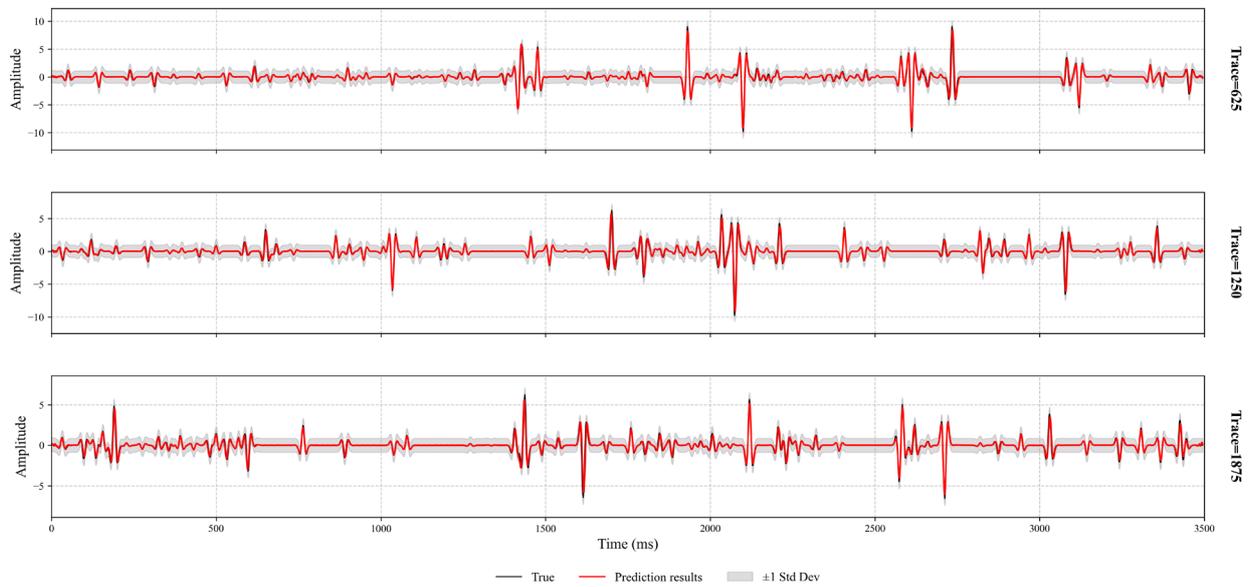

Figure 6. Comparison of predicted and true seismic data traces at selected locations (625, 1250, 1875) along the horizontal axis.

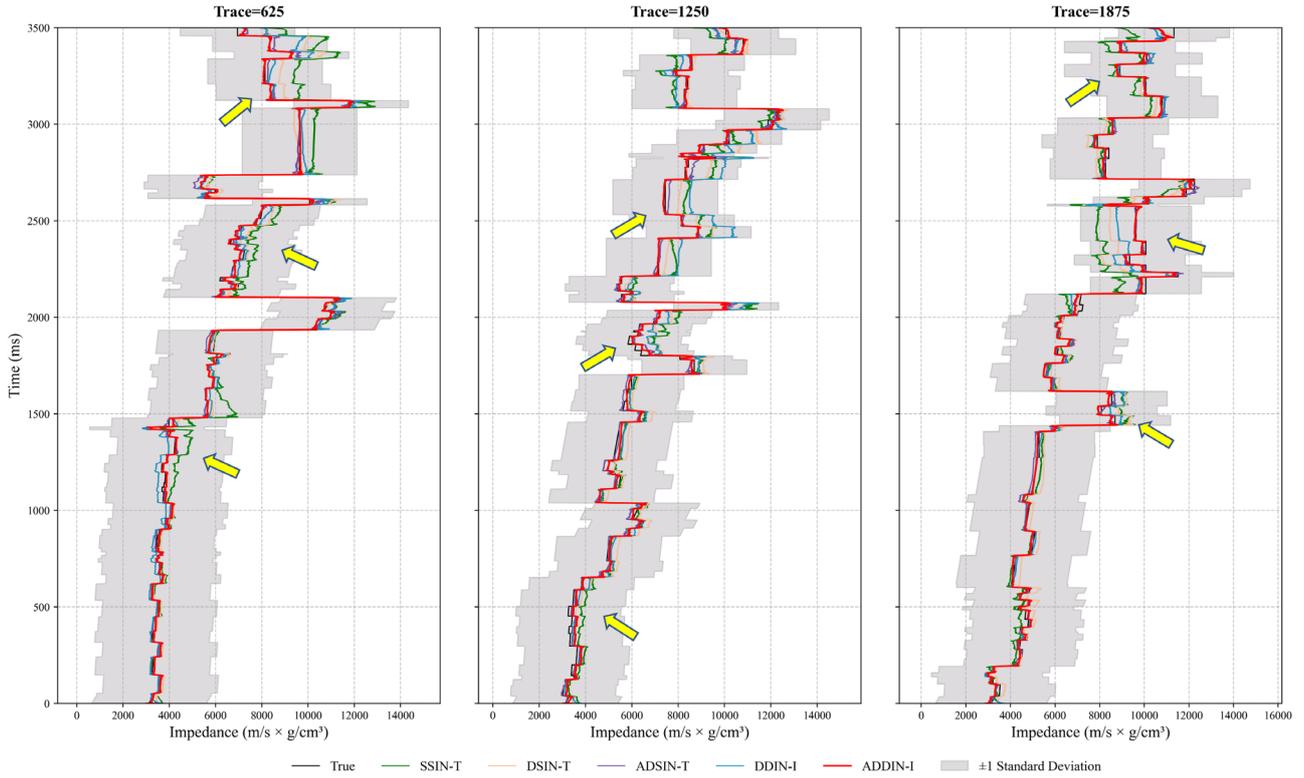

Figure 7. Comparison of predicted and true acoustic impedance traces at selected locations (625, 1250, 1875) along the horizontal axis.

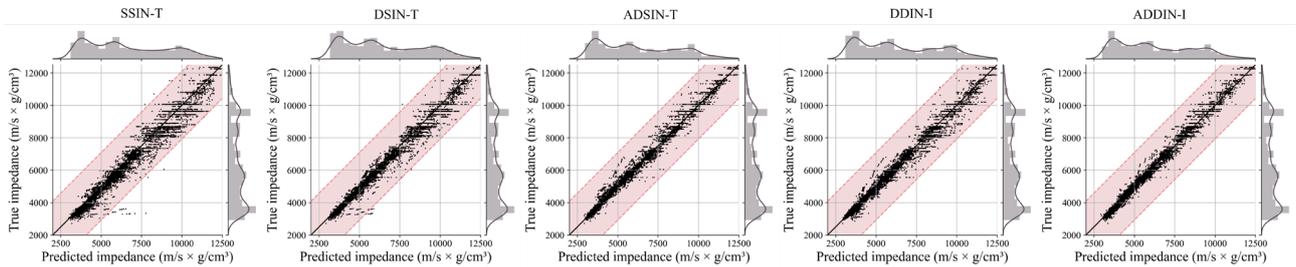

Figure 8. Scatter plots of different methods.

Figure 9 displays the convergence of loss functions for five methods: (a) training loss curves and (b) validation loss curves. The ADDIN-I method (blue curve) converges after about 200 iterations, exhibiting faster convergence and lower loss values than other methods. This rapid convergence underscores the model's high accuracy. The ADDIN-I and ADSIN-T methods (blue and cyan curves) demonstrate notably stable loss curves. This stability can be attributed to the attention mechanism's ability to dynamically focus on critical input data, enhancing the model's adaptability to data variations. Consequently, this approach not only stabilizes loss curves but also enhances the model's robustness and generalization capabilities.

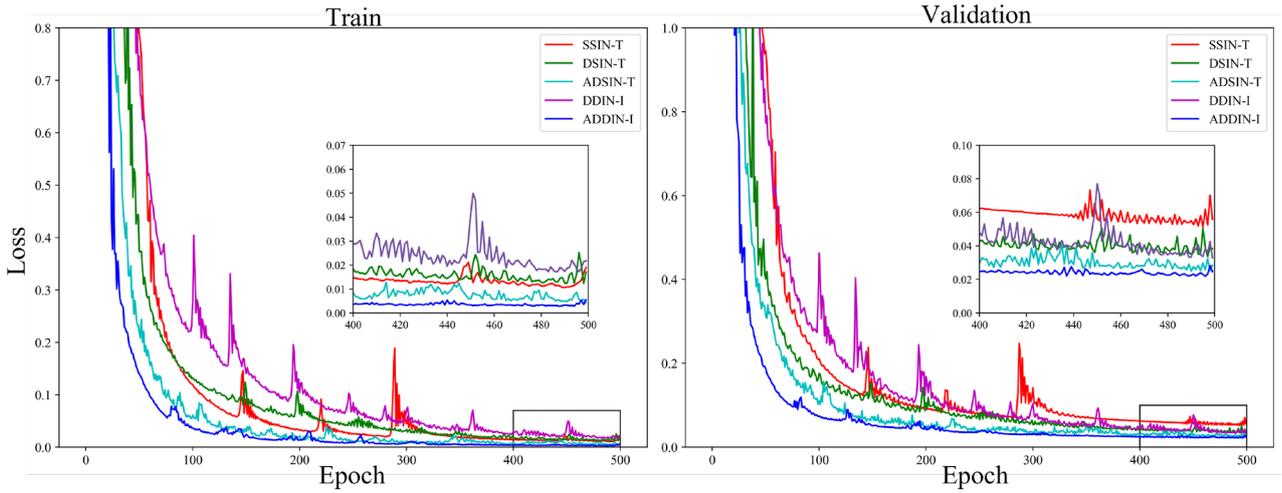

Figure 9. Training and validation loss curves for five methods. (a) Training loss; (b) Validation loss.

## 3.4 Field data application

Following its successful application to synthetic model data, we applied the ADDIN-I method to real data for further validation. This study utilized real seismic and logging data from the South China Sea region. The region's complex lithology, significant thickness variations, and strong heterogeneity present substantial challenges for seismic impedance inversion. Figure 10 illustrates the regional profile. The experimental well-to-well profile comprises 536 seismic traces, each with approximately 360 samples at a 1 ms depth sampling rate. The study incorporated data from wells H1, H2, and H3 alongside the seismic traces. Wells H1 and H2 served as the training dataset, with well H3 used for validating the inversion results.

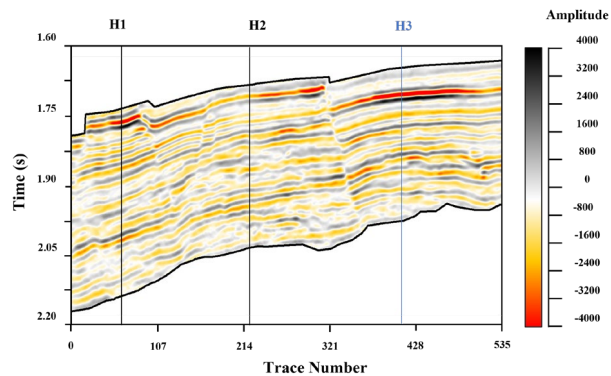

Figure 10. Well-tied seismic profile.

The forward modeling network directly generates synthetic seismic records from real seismic data, eliminating the need to extract actual seismic wavelets. This approach enables effective semi-supervised training, showcasing deep learning's advantages in inversion tasks. Model data has validated the effectiveness of each module. In real data applications, we compared the dual-branch network with attention mechanisms to the single-branch network to demonstrate its advantages. We conducted comparative experiments using two methods with identical parameter modules: the single-branch double-inversion network without attention mechanisms (SDIN-I) and the dual-branch double-inversion network without attention mechanisms (DDIN-I). To further demonstrate the superiority of deep learning methods, we also compared our approach with the traditional Bayesian Model Inversion Method (BMI) using real data.

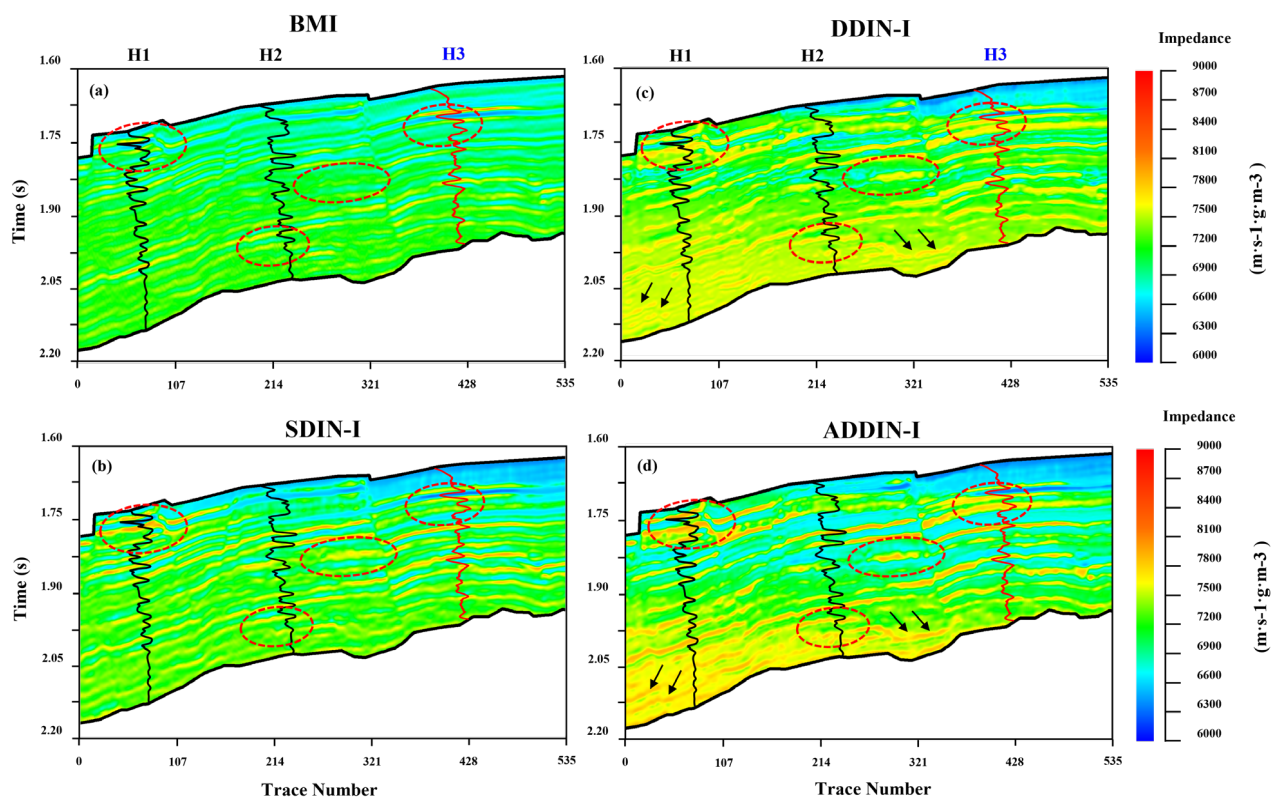

Figure 11. Inversion results for actual seismic data using different methods. (a) Bayesian Model Inversion; (b) SDIN-I; (c) DDIN-I; (d) ADDIN-I.

Figure 11 presents the inversion results for BMI, SDIN-I, DDIN-I, and ADDIN-I methods. Well logs from H1, H2, and H3 are overlaid on the inversion results for comparison. SDIN-I, DDIN-I, and ADDIN-I inversion results closely match the H3 well log. The BMI method shows good lateral continuity but poor vertical resolution, resulting in mismatches with the well log in high-impedance areas (red box, H3, Figure 11a). This limitation hinders accurate identification of thin layers and faults. In the red boxes near H1 and H2, ADDIN-I demonstrates superior vertical resolution, more continuous lateral features, and clearer stratigraphic structures compared to SDIN-I and BMI. This highlights the advantage of the dual-branch structure in feature extraction. Between H2 and H3 (red box), ADDIN-I reveals more details and better lateral continuity. The black arrows in Figures 11c and 11d show that ADDIN-I highlights more stratigraphic impedance details, providing more accurate information. This showcases the positive role of the attention mechanism. These results align with the numerical model experiments, further validating ADDIN-I's effectiveness and practicality. Figure 12 compares actual and predicted wave impedance curves for well H3 (trace 410). The analysis shows ADDIN-I yields higher prediction accuracy and stronger correlation with the actual wave impedance curve. Table 3 presents Pearson correlation coefficients between the H3 well log and inversion results from BMI, SDIN-I, and ADDIN-I. In conclusion, numerical experiments demonstrate ADDIN-I's superior performance over comparison methods.

Table 3. Pearson Correlation Coefficients (PCC) between H3 Well Data and Inversion Results for the 410th Trace

| Method | PCC |
| --- | --- |
| BMI | 0.6543 |
| SDIN-I | 0.8114 |
| DDIN-I | 0.8664 |
| ADDIN-I | **0.8827** |

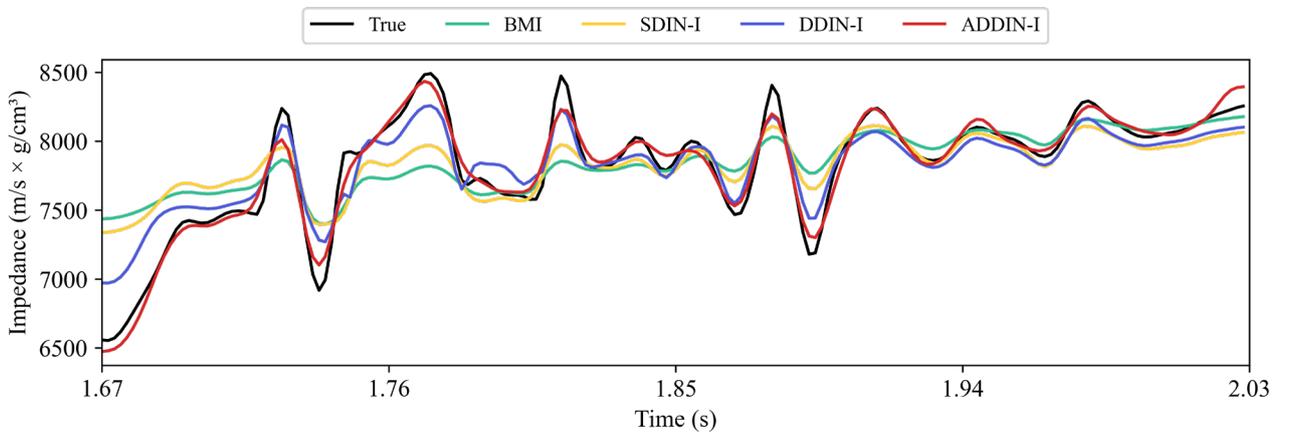

Figure 12: Comparison of inversion results for trace 410 using different methods against H3 well data.

## 4. Conclusions

We propose a dual-branch double-inversion network method for seismic impedance inversion based on attention mechanisms (ADDIN-I), which enhances the accuracy of inversion results. First, this method employs a branching network strategy to effectively address the difficulty of extracting high-frequency weak signal features with a single network. The combination of Bi-GRU and TCN networks strengthens the correlation between data points, allowing the network to capture the long-term dependencies in seismic data, thereby achieving high-precision predictions. The integration of attention mechanisms further improves the prediction accuracy and stability of network training. Additionally, an improved semi-supervised network training strategy increases the weight of near-well data during the training process, enhancing overall performance. By incorporating a deep learning forward operator, the network can be applied to real seismic data. The application of this method to model data for impedance inversion demonstrated the effectiveness of each module. The method was successfully applied to actual seismic data, yielding inversion results superior to those of traditional Bayesian Model Inversion (BMI), the single-branch dual-inversion network method without attention mechanisms (SDIN-I), and the dual-branch double-inversion network method without attention mechanisms (DDIN-I). The combined results of model experiments and real data applications validate the effectiveness of the proposed method. However, despite the improved semi-supervised network training strategy alleviating the dependency on the training dataset to some extent, this method remains entirely data-driven. Future work will involve introducing a known physical system to establish a label-free closed-loop data-driven framework and capturing the uncertainty of inversion results to achieve reliability assessments of the inversion results.

109549.